\documentclass[12pt]{article}
\usepackage{graphicx}
\usepackage{comment}
\RequirePackage{color}
\RequirePackage{mathtools}
\RequirePackage{snapshot}
\RequirePackage[english]{babel}
\RequirePackage[utf8]{inputenc}
\RequirePackage{cuted}
\RequirePackage{booktabs}
\RequirePackage{multirow}
\RequirePackage{url}
\RequirePackage[numbers,sort&compress]{natbib}
\RequirePackage[normalem]{ulem}
\RequirePackage{changepage}

\newcommand\pubnumber{}
\newcommand\pubdate{\today}

\def\institute{Institute for Theoretical Particle Physics and Cosmology\\
RWTH Aachen University \\ D-52056 Aachen, Germany}
\def\support{\footnote{Work supported by German Research Foundation (DFG) Collaborative Research Centre/Transregio project CRC/TRR 257:P3H - \textit{Particle Physics Phenomenology after the Higgs Discovery}, by the Research Training Group GRK $2497$: \textit{The physics of the heaviest particles at the Large Hadron Collider} and by a grant of the Bundesministerium f\"ur Bildung und Forschung (BMBF).  }}

\def\Title#1{\begin{center} {\Large #1 } \end{center}}
\def\Author#1{\begin{center}{ \sc #1} \end{center}}
\def\Address#1{\begin{center}{ \it #1} \end{center}}

\newcommand{\PrePrintNumbersname}{{Preprint numbers}}
\newcommand\pubblock{\rightline{\begin{tabular}{l} \pubnumber\\
         \pubdate  \end{tabular}}}
\newenvironment{Abstract}{\begin{quotation}  }{\end{quotation}}
\newenvironment{Presented}{\begin{quotation} \begin{center} 
             PRESENTED AT\end{center}\bigskip 
      \begin{center}\begin{large}}{\end{large}\end{center} \end{quotation}}

\def\beq{\begin{equation}}
\def\eeq#1{\label{#1}\end{equation}}
\def\eeqn{\end{equation}}

\def\beqa{\begin{eqnarray}}
\def\eeqa#1{\label{#1}\end{eqnarray}}
\def\eeqan{\end{eqnarray}}

\let\bar=\overbar

\def\Dslash{\not{\hbox{\kern-4pt $D$}}}
\def\dslash{\not{\hbox{\kern-2pt $\del$}}}

\def\msb{{\bar{\ssstyle M \kern -1pt S}}}

\begin{document}
\begin{titlepage}
\pubblock

\vfill
\Title{On the exclusion limits in $t\bar{t}+$DM searches at the LHC}
\vfill
\Author{ Jonathan Hermann\support}
\Address{\institute}
\vfill
\begin{Abstract}
In these proceedings we discuss the importance of higher-order corrections and off-shell effects for the calculation of signal strength exclusion limits in $t\bar{t}+$DM searches at the Large Hadron Collider. We present limits for the spin-0 s-channel mediator model using state-of-the-art NLO QCD predictions with full off-shell effects for the relevant $t\bar{t}$ and $t\bar{t}Z$ background processes. These are then compared to less sophisticated predictions using either the narrow-width approximation or LO calculations in order to asses the respective effects.
\end{Abstract}
\vfill
\begin{Presented}
$14^\mathrm{th}$ International Workshop on Top Quark Physics\\
Michigan, US (videoconference), 13--17 September, 2021
\end{Presented}
\vfill
\begin{center}
	\PrePrintNumbersname \\ TTK-22-03 $\,\cdot\,$ P3H-22-003
\end{center}
\end{titlepage}
\def\thefootnote{\fnsymbol{footnote}}
\setcounter{footnote}{0}

\section{Introduction}
Even though the existence of Dark Matter (DM) has been well established, the discovery of a suitable Dark Matter candidate has proven to be a difficult task.
At the forefront of current DM searches are collider searches by the CMS 
and ATLAS 
collaborations. 
These analyses cover a wide range DM models which result in various signatures and final states. 
The unifying characteristic of all of these models is that they result in an excess of missing transverse momentum  $p_{T,\text{miss}}$ which might be observable at the LHC. 
However, in order to distinguish any sort of signal from the SM background, one requires accurate and precise predictions for the Standard Model (SM) background. 
This is particularly true for distribution tails, e.g. for $p_{T,\text{miss}}$, as both next-to-leading (NLO) QCD corrections and off-shell effects are most prominent in these tails.
NLO QCD corrections have by now become the norm for (associated) top-quark pair production but decays are often still modelled at leading order (LO), like e.g. in \textsc{MadGraph5\_aMC@NLO} \cite{MadGraph}, and off-shell effects are often ignored or only partly taken into account.
As full off-shell NLO QCD predictions have become available over past few years (see e.g. Refs \cite{tt,ttZ}), we want to asses the importance of both higher-order corrections and off-shell effects and evaluate whether the added complexity is actually necessary for DM searches at the LHC. The discussion is based on the results presented in Ref. \cite{tt_DM_paper}.


\section{Setup}

For our analysis we consider top-quark pair associated DM production with leptonic top decays, i.e. $p p \to b \bar{b} e^+ \mu^- \nu_e \bar{\nu}_{\mu} \chi \bar{\chi}$ production. The DM particle pair production is described by the spin-0 s-channel mediator model which extends the SM by a (pseudo-)scalar mediator $Y$ and a fermionc DM particle $\chi$. Adhering to the Minimal Flavor Violation (MFV) hypothesis, the coupling between the mediator and SM quarks is proportional to the Yukawa coupling $y_q$. 
We fix the DM particle mass to $m_\chi = 1$ TeV and the couplings to $g_q = g_\chi = 1$, in line with the recommendations outlined in Ref. \cite{DM_simp_1}. 
The mediator mass is varied between $m_Y = 10$ GeV and $1$ TeV.
The $t \bar{t} \chi \bar{\chi}$ signal predictions are calculated with \textsc{MadGraph5\_aMC@NLO} at NLO in QCD using the \textsc{DMsimp} \cite{DM_model_Backovic} implementation.

The dominant SM background processes for the $t\bar{t}+$DM signal are $t\bar{t}$ and $t\bar{t}Z,\,Z\to\nu\bar{\nu}$ production with leptonically decaying top quarks. Cross-sections and distributions for these processes are calculated using \textsc{Helac-NLO} \cite{HELAC_NLO, HELAC_NWA}. Integrated fiducial cross-sections with inclusive cuts are presented in Table \ref{table:total_xSec_NLO_cut} as $\sigma_{\text{uncut}}$.

\begin{table}[t]
	\caption{\textit{Comparison of LO and NLO integrated cross sections for the two background processes in the NWA (top) and the FOST (bottom) before and after applying the additional cuts. The numbers of events are given for an integrated luminosity of $L = 300 \,\text{ fb}^{-1}$ and include the lepton flavour factors ($4$ for the DM signal and $t\bar{t}$, and $12$ for $t\bar{t}Z$).}}
	\centering
		\begin{tabular}{ll@{\hskip 10mm}lll@{\hskip 10mm}l}
			\hline\noalign{\smallskip}
			Process  & Order & $\sigma_{\text{uncut}}$ [fb] & $\sigma_{\text{cut}}$ [fb] & $\sigma_{\text{cut}} / \sigma_{\text{uncut}}$ & Events\\
			\noalign{\smallskip}\midrule[0.5mm]\noalign{\smallskip}
			\multirow{3}{*}{$t\bar{t}$ NWA}  & LO              & $1061$    & $ 0 $    & $ 0.0 \%$  & $ 0 $\\
			& NLO             & $1097 $   & $ 0 $    & $ 0.0 \%$    & $ 0 $\\
			& NLO with LO decays &  $1271 $   & $ 0 $    & $ 0.0 \%$    & $ 0 $\\
			\noalign{\smallskip}\hline\noalign{\smallskip}
			\multirow{3}{*}{$t\bar{t}Z$ NWA}  & LO          & $0.1223 $ & $0.0130 $               & $ 11\%$     & $ 47$\\
			& NLO         & $ 0.1226 $  & $ 0.0130 $               & $ 11 \%$    & $ 47$\\
			& NLO with LO decays  & $ 0.1364 $  & $ 0.0140 $               & $ 10 \%$    & $ 50 $\\
			\noalign{\smallskip}\hline\noalign{\smallskip}
			\multirow{2}{*}{$t\bar{t}$ Off-shell}       & LO   & $1067$ & $0.0144$ & $0.0013 \%$ & $17$\\
			& NLO        & $1101 $ & $0.0156$ & $0.0014 \%$ & $19$\\
			\noalign{\smallskip}\hline\noalign{\smallskip}
			\multirow{2}{*}{$t\bar{t}Z$ Off-shell}  & LO   & $0.1262 $ & $0.0135$ & $11 \%$ & $49$\\
			& NLO       & $0.1269$ & $0.0134$ & $11 \%$ & $48$\\
			\noalign{\smallskip}\hline
		\end{tabular}
		\label{table:total_xSec_NLO_cut}
\end{table}

For the actual analysis we impose an additional, more exclusive cut selection based on Ref. \cite{Haisch_analysis}. Here, we also require $p_{T,\text{miss}} > 150$ GeV, $M_{T2,W} > 90$ GeV and $C_{em,W} > 130$ GeV where $M_{T2,W}$ is the stransverse $W$-mass and $C_{em,W}$ is defined as $C_{em,W} =  M_{T2,W} - 0.2\cdot (200 \text{ GeV} - p_{T,\text{miss}})$. 
The corresponding integrated fiducial cross-sections are listed in Table \ref{table:total_xSec_NLO_cut} under $\sigma_{\text{cut}}$ along with the fraction of events that pass the cuts and the expected number of background events for an integrated luminosity of $L=300 \,\text{fb}^{-1}$.

It turns out that despite the initially overwhelming size of the $t\bar{t}$ cross-section, the $t\bar{t}Z$ process is actually the dominant SM background with between $47$ and $50$ expected events depending on the modeling, compared to up to $19$ for $t\bar{t}$. However, the most important observation is that in the narrow-width approximation (NWA), no $t\bar{t}$ events pass the exclusive cuts which is not the case in the full off-shell treatment (FOST). This is a result of the cut on $M_{T2,W}$ since for $t\bar{t}$ in the NWA, we have $M_{T2,W} \leq m_{W}$. However, single- and non-resonant contributions still appear above this threshold so that in the FOST there are still $t\bar{t}$ events passing the cuts. This means that around a quarter of the background events is missing when using the NWA instead of the FOST which significantly affects the exclusion limits.

\begin{figure*}
	\includegraphics[width=.5\linewidth]{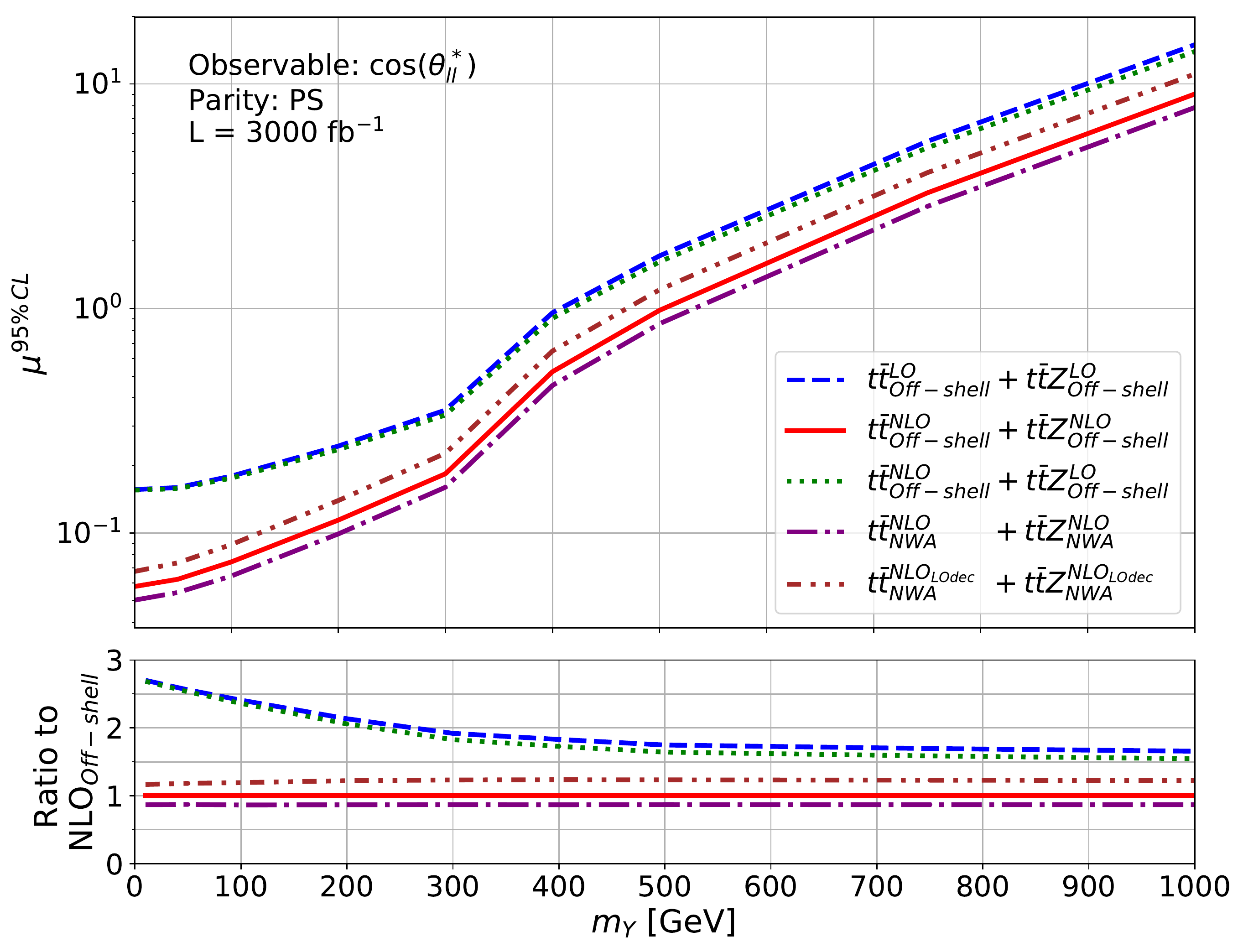}
	\includegraphics[width=.5\linewidth]{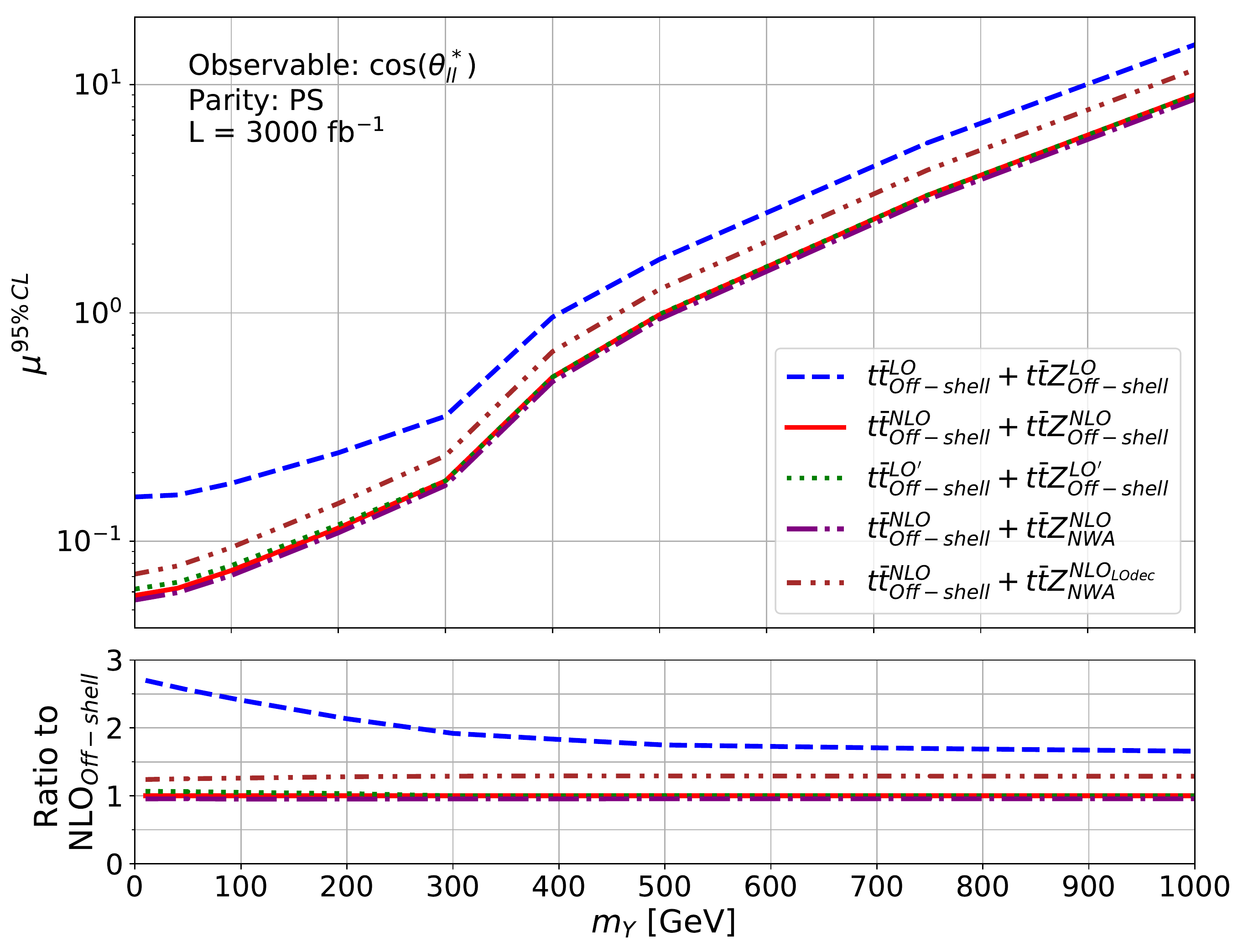}
	
	\caption{\textit{Comparison of signal strength exclusion limits computed with different background predictions for the pseudoscalar mediator scenarios with a luminosity of $L = 3000 \text{ fb}^{-1}$. In the lower panels we present the ratios to the limits obtained using the NLO$_{\text{Off-shell}}$ background predictions.}}
	\label{fig:Limits_Bkg_comp}
\end{figure*}


\section{Exclusion Limits}
In Figure \ref{fig:Limits_Bkg_comp} we compare the signal strength exclusion limits $\mu^{95 \%\,\text{CL}}$ calculated at $95 \%$ confidence level (CL) for various different SM background predictions. All limits were computed using the \textsc{HistFitter} \cite{HistFitter} implementation of the $CL_s$-method \cite{CLs_ALRead}. For this specific comparison we chose $\cos ( \theta^*_{ll}) = \tanh(|\eta_{l_1} - \eta_{l_2}|/2)$ as our observable. An explicit comparison between the integrated fiducial cross-section $\sigma_{\text{tot}}$, $p_{T,\text{miss}}$, $\cos ( \theta^*_{ll})$, the azimuthal angle between the missing transverse momentum and the closest lepton $\Delta \phi_{l,\text{miss}}$ and the stransverse top mass $M_{T2,t}$ showed that $\cos ( \theta^*_{ll})$ yields the most stringent limits for almost all considered mediators. Solely for light scalar mediators, $M_{T2,t}$ provides better limits.
The results shown here were obtained using the pseudoscalar mediator scenario but they are comparable for the scalar case. We use an integrated luminnosity of $L = 3000 \,\text{fb}^{-1}$ for this specific comparison but the main observations are independent of this value. However, the differences are more striking for larger luminosities as statistical uncertainties play less of a role.

One can clearly observe from the left plot in Figure \ref{fig:Limits_Bkg_comp} that the off-shell NLO (red) background predictions lead to significantly better limits than if one uses LO predictions (blue). We also find that it is insufficient to only model the $t\bar{t}$ process at NLO as the $t\bar{t}^{NLO} + t\bar{t}Z^{LO}$ background (green) leads to limits that are almost as far off the full NLO results as the pure LO results. This is not surprising as we have already seen that $t\bar{t}Z$ is the dominant background, so reducing the scale uncertainties is more important for this process than for $t\bar{t}$.

Turning to the off-shell effects, we observe that the NWA predictions lead to underestimated limits (purple) compared to the full prediction. The main reason are the missing $t\bar{t}$ events in the NWA which we mentioned previously. This hypothesis is underlined by the comparison on the right side of Figure \ref{fig:Limits_Bkg_comp} where we replaced the $t\bar{t}$ NWA predictions by their off-shell counterpart whilst keeping $t\bar{t}Z$ in the NWA. As this substantially reduces the difference to the NLO off-shell limits, we conclude that off-shell effects in $t\bar{t}Z$ only play a minor role here.

We also considered NLO predictions with LO decays (brown) in this comparison. Since the size of their scale uncertainties falls in between those for the LO and NLO predictions, the exclusion limits do the same. As these limits are considerably worse than those for the full NLO, we can conclude that it is important to also include NLO corrections to the decays.

In order to disentangle the effects of reduced scale uncertainties and shape distortions between LO and NLO, we also include a so-called LO' prediction on the right plot (green) which corresponds to LO predictions with NLO uncertainites. The corresponding limits essentially coincide with the full NLO results, so the main difference between LO and NLO is the reduction in scale uncertainties while shape distortions only play a minor role. Let us mention that in principle the integrated cross-section can also change significantly between LO and NLO but due to our scale choice this effect is minimized.

In addition to the above described comparison, we also considered the effect that the choice of the central scale has on the exclusion limits. It turns out that for some observables, in particular $M_{T2,t}$, this is even relevant at NLO. As a fixed scale setting leads to larger scale uncertainties than the dynamical ones, the corresponding limits are also worse. Hence, it is important to use a dynamical scale. In contrast, the difference between the tested dynamical scale settings is rather small.

We also investigated the effects of different integrated luminosities on the exclusion limits. As one might expect, the limits improve when increasing the luminosity since statistical uncertainties are proportional to $1 / \sqrt{L}$. These effects are larger the more precise the background prediction is because statistical uncertainties make up a larger proportion of the total uncertainty for more precise backgrounds. 

For further details on all of the described comparisons we refer to Ref. \cite{tt_DM_paper}.

\section{Conclusions}
In these proceedings we have discussed various different inputs for the calculation of exclusion limits and outlined the impact that changing these has on the limits. We have demonstrated that NLO QCD corrections are vital for this type of analysis. Off-shell effects also need to be taken into account for the $t\bar{t}$ process but seem to be unnecessary for $t\bar{t}Z$. In addition, we found that the central scale choice is important even at NLO and that one should refrain from using a fixed scale. Similar to higher-order corrections, an increased luminosity reduces the uncertainties and thus also improves the limits considerably. Finally, we investigated which observable yields the most stringent limits and identified $\cos ( \theta^*_{ll})$ as the most promising observable.


\end{document}